\documentclass[11pt]{article}

\usepackage[final]{acl}

\usepackage{times}
\usepackage{latexsym}

\usepackage[T1]{fontenc}

\usepackage[utf8]{inputenc}

\usepackage{microtype}

\usepackage{inconsolata}

\title{Where Did It Go Wrong? Capability-Oriented Failure Attribution for Vision-and-Language Navigation Agents}

\author{
 \textbf{Jianming Chen\textsuperscript{1,2,3,4}},
 \textbf{Yawen Wang\textsuperscript{1,2,3,4}}\thanks{Corresponding authors.},
 \textbf{Junjie Wang\textsuperscript{1,2,3,4}}\footnotemark[1],
 \textbf{Xiaofei Xie\textsuperscript{5}},
 \textbf{Shoubin Li\textsuperscript{1,2,3,4}},
\\
 \textbf{Qing Wang\textsuperscript{1,2,3,4}},
 \textbf{Fanjiang Xu\textsuperscript{1,2,3,4}}\footnotemark[1]
\\
\textsuperscript{1}Institute of Software, Chinese Academy of Sciences, Beijing, China
\\
\textsuperscript{2}Science \& Technology on Integrated Information System Laboratory, Beijing, China
\\
\textsuperscript{3}State Key Laboratory of Complex System Modeling and Simulation Technology, Beijing, China
\\
\textsuperscript{4}University of Chinese Academy of Sciences, Beijing, China
\\
\textsuperscript{5}Singapore Management University, Singapore
\\
\small{
    \{jianming2023, yawen2018, junjie, shoubin, wq, fanjiang\}@iscas.ac.cn, xfxie@smu.edu.sg
}
}

\usepackage{amssymb}
\usepackage{multirow}
\usepackage{tcolorbox}
\usepackage{soul}
\usepackage{makecell}
\usepackage{mathtools}
\usepackage[ruled, linesnumbered]{algorithm2e}
\usepackage{booktabs}
\usepackage[table]{xcolor}
\usepackage{enumitem}

\newcommand{\tool}{CanTest}
\newcommand{\website}{https://github.com/JMChen121/CanTest/}

\begin{document}
\maketitle

\begin{abstract}
Embodied agents in safety-critical applications such as Vision-Language Navigation (VLN) rely on multiple interdependent capabilities (e.g., perception, memory, planning, decision), making failures difficult to localize and attribute. Existing testing methods are largely system-level and provide limited insight into which capability deficiencies cause task failures. We propose a capability-oriented testing approach that enables failure detection and attribution by combining (1) adaptive test case generation via seed selection and mutation, (2) capability oracles for identifying capability-specific errors, and (3) a feedback mechanism that attributes failures to capabilities and guides further test generation. Experiments show that our method discovers more failure cases and more accurately pinpoints capability-level deficiencies than state-of-the-art baselines, providing more interpretable and actionable guidance for improving embodied agents.

\end{abstract}

\section{Introduction}
\label{sec-intro}

Recently, embodied agents have garnered increasing attention for their potential to interact with various environments in applications such as navigation assistance for the visually impaired \cite{vp1,vp2,vp3} and home service robots \cite{robot1,robot2,robot3,robot4}.
These agents are designed to perform tasks that demand safety and efficiency, making their reliability essential in daily life.
Any errors in their operations can lead to severe consequences, such as visually impaired users being misdirected from safe routes \cite{vp4}, or home service robots neglecting to turn off appliances, which could result in safety hazards \cite{robot2}.
Therefore, comprehensive testing to ensure these agents work reliably under diverse conditions is crucial.

Embodied agents integrate and rely on a combination of capabilities to perform complex tasks \cite{vln1}, including perception, memory, planning, and decision, with each capability serving as a component within the agent.
However, due to the close interdependence of these capabilities, deficiencies in one often cascade to others, leading to compounded errors.
This interplay of errors can be particularly pronounced in tasks that involve lengthy sequences, as mistakes can propagate and compound over time.
A prominent illustration of this phenomenon can be found in Vision-Language Navigation (VLN) agents, which are a representative type of embodied agent designed to follow natural language instructions and navigate through environments \cite{vln2}.
In VLN tasks, failing to associate a visual landmark with the corresponding instruction (a perception error) could lead to incorrect memory updates, flawed planning, and ultimately poor decision-making during navigation \cite{vln_Perception,vln_memory,vln_plan1}.


Although existing test case generation techniques are effective in identifying system-level failures, they fall short for embodied agents due to the interplay and error propagation among capabilities \cite{fuzz,MoDitector,BehAVExplor}.
Traditional methods treat agent as a monolithic entity, detecting task failures without revealing which capability caused the error or how it emerged \cite{attack}.
The absence of capability-oriented failure attribution restricts developers from precisely locating and resolving weaknesses, hindering targeted improvements.
Therefore, there is an urgent need for a capability-oriented testing approach tailored to embodied agents—one that generates test cases enabling precise failure localization at capability level. By isolating issues within specific capabilities, it would offer greater interpretability and more actionable insights for developers.



To enable capability-oriented testing of embodied agents, two key challenges must be addressed.
\textbf{First}, it is necessary to construct capability-oriented test oracles that assess whether capability outputs are correct.
Automatically building such oracles is difficult, as it requires anticipating diverse scenarios and designing distinct evaluation metrics for different capabilities.
\textbf{Second}, effective failure attribution over long task trajectories is challenging.
Each time step in a sequence involves a chain with multiple capabilities interacting, and these steps collectively form an even longer chain of task execution.
Errors introduced by one capability often propagate and compound over time, making it difficult to trace back to the initial source of failure. 
Developing a mechanism to identify the initial error along this chain is essential.

In this paper, we propose a novel \textbf{\textit{Ca}}pability-orie\textbf{\textit{n}}ted \textbf{\textit{Test}}ing approach, {\tool}, for revealing and attributing failures to specific capabilities.
Our approach includes: (1) a process for automatic generation of test cases, with the adaptive selection and mutation of case seeds, to generate test cases that are likely to expose capability-oriented failures.
More notably, to address the first challenge, we introduce (2) the construction of capability oracles, which extract expected outputs and define independent evaluation metrics for each capability to determine whether errors occur.
To address the second challenge, we design (3) a feedback mechanism that leverages the oracles to identify the capability responsible for a failure and computes a feedback score that considers both the failure and its source capability error.
By integrating failure-oriented and capability-oriented measurements, {\tool} computes feedback scores that guide the generation of test cases designed to expose both capability deficiencies and task-level failures.

To evaluate the effectiveness of {\tool}, we take the VLN task as the subject of our study, and conduct experiments targeting three advanced VLN models, with results indicating that {\tool} is capable of discovering the largest number of failure cases and outperforms three baselines, which are either commonly used or state-of-the-art (SOTA).
The improvement in the number of discovered failure cases relative to the best-performing baseline ranges from 23.34\% to 33.70\%.
Moreover, we conduct an experiment that plugs the capability oracles and attribution mechanisms constructed by {\tool} into the baseline, expanding their ability to provide failure attribution. However, the results indicate that {\tool} is still capable of discovering the largest number of failure cases targeting each capability.
Next, we repair the failure case using our capability oracles, achieving repair rates ranging from 81.30\% to 96.69\%, thereby demonstrating the high fidelity of the oracles.
Finally, through ablation experiments, we validate the significance of the feedback scores we design, showing that both failure-oriented and capability-oriented feedback contribute to discovering more failure cases.
The main contributions of this work are as follows:

\begin{itemize}[noitemsep]

    \item To the best of our knowledge, this is the first work to propose automated test case generation specifically targeted at the individual capabilities of embodied agents.
    
    \item We automatically construct a novel set of capability-oriented test oracles designed to independently evaluate each capability of embodied agents, e.g., the perception, memory, planning, and decision-making capabilities within VLN agents.
    
    \item We designed a feedback mechanism that attributes task failures to the specific capability error and thus provides the feedback scores that comprehensively account for both failure and capability error.
    
    \item 
    Experimental evaluations on the effectiveness of {\tool} outperform all baselines, with promising performance in terms of the number of failure cases, which can all be attributed to specific capability errors\footnote{\url{\website}}. 
    
\end{itemize}

\section{Related Work}
\label{sec:related}

\subsection{Testing VLN Agent}

Traditionally, the quality assessment of embodied agents has relied heavily on task-specific metrics designed to quantify performance in controlled environments \cite{VLAevaluating}. For instance, in the realm of navigation and path planning, evaluations are often conducted using various metrics, including path length, execution time, and energy consumption \cite{vln1, vln2}.

In addition, many of these tests \cite{Palm-e, RT2, VLATest,LADEV} are primarily designed for simulation-based robots, providing various scenarios for a diverse range of tasks, such as manipulation and navigation. 
They often focus on scene setup and task completion without capturing the full range of capabilities required by the agents, such as reasoning quality and task performance quality.
For instance, in the benchmarks used to evaluate PaLM-E \cite{Palm-e} and RT-2 \cite{RT2}, the authors emphasize the final task performance of specific agents in complex environments.
VLATest \cite{VLATest} is a fuzzy framework designed to generate robotic operational scenarios for testing agents, which also uses a simple oracle to evaluate the correctness of task completion.


\subsection{Technology of Test Case Generation}

Existing approaches primarily employ the following technologies to implement automatic test case generation: 
Evolutionary Algorithms \cite{EA1,EA2}, which are inspired by biological evolution principles to optimize test case generation. For example, AVUnit \cite{EA2} supports various fuzz testing algorithms that use robustness and coverage as fitness metrics to automatically search for test cases that violate these assertions.
Model-Based Search \cite{modelsearch1,modelsearch2,modelsearch3}, where algorithms utilize models of the system under test to guide the search for valid test cases. For instance, SAMOTA \cite{modelsearch2} extends existing multi-objective search algorithms for test case generation, enabling efficient use of alternative models with lower running costs.
Fuzzing Methods \cite{fuzz,MoDitector,BehAVExplor}, which involve sending random or semi-random inputs to uncover vulnerabilities. For example, MoDitector \cite{MoDitector} can effectively generate conflict or failure scenarios and can also establish relationships between module errors and system failures.

\begin{figure*}[tbp]
    \centering
    \includegraphics[width=\textwidth]{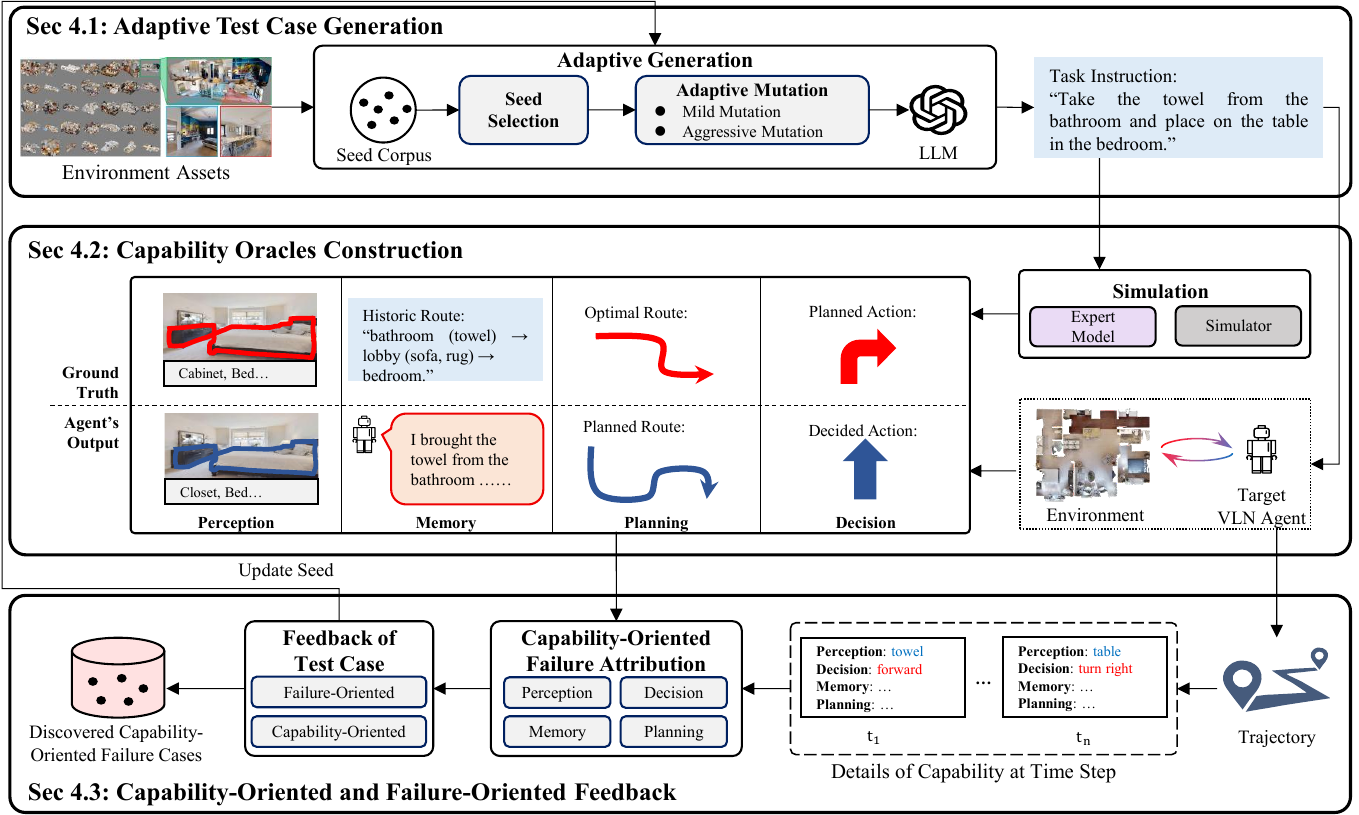}
    \caption{
    Overview of {\tool}.
    }
    \label{fig:overview}
\end{figure*}

\section{Problem Statement}
\label{sec:background}

\subsection{Vision-Language Navigation}
The VLN agents aim to enable intelligent agents to perform navigation tasks in unknown environments based on natural language instructions \cite{vln_plan1}.
The agent receives a natural language instruction $I$ as input (e.g., "Walk from the kitchen to the living room, then enter the green door leading to the bedroom").
The instruction is represented as a word sequence: $I = \{w_1, w_2, w_3, \dots, w_L\}$,
where $w_i$ denotes the $i$-th word in the instruction, and $L$ is the total number of words. 

The agent perceives the environment from an egocentric view and acts through viewpoint changes or low-level actions (e.g., move forward, turn left/right), producing a trajectory $\tau={s_1,\dots,s_T}$ where $s_t$ denotes state. VLN remains challenging due to ambiguous instructions and environmental uncertainty, e.g., dynamics and noisy visual inputs \cite{Fine-grained}.

\subsection{\textbf{Failures of VLN Task}}
\label{sec:failure}
In the VLN task, failures of the agent are typically defined as the inability of the agent to successfully complete the given navigation task.
Specifically, failures can be determined from the aspects of \textbf{target position} and \textbf{step limit}\cite{vln_state}, as shown in Appendix \ref{appx:background}.


Regardless of whether the endpoint is too far from the target position or whether it has consumed additional time steps outside the limit, such situations will be deemed as task failures. In this paper, we use "\textit{error}" to refer to the deviation between the capability output and the expected output, while "\textit{failure}" signifies that the task has not been completed. Furthermore, it is worth noting that not all capability errors will lead to task failures.



\subsection{Problem Statement}
\label{sec:problem}

We aim to develop a capability-oriented testing approach for embodied agents.
An automated test case generation mechanism is required to produce the test case set $TC = \{tc_1, tc_2, \dots, tc_n\}$ that explores the case space comprehensively.
Let a target agent $A = \{C_p, C_m, C_{pl}, C_d\}$ integrating multiple capabilities, where $C_p$, $C_m$, $C_{pl}$, and $C_d$ denote the capability of perception, memory, planning, and decision, respectively.
The approach should enable the construction of test oracles $\{TO_p, TO_m, TO_{pl}, TO_d\}$ for each capability $C_x \in A$ to evaluate their quality and leverage these oracles to diagnose capability-oriented errors in the trajectory $\tau$. 

Finally, 
for the test case $tc_i$, if it is a failure case, we can attribute the failure to a specific capability:
\begin{equation}
    \label{equation:problem}
    \begin{aligned}
    & (\forall \tau_i = A(tc_i), tc_i\ is \ failure)        \\
    & \land  ( C^{errors}=\{C_x|TO_x(C_x)=False\} )         \\
    & \Rightarrow \exists !C^*_x \in C^{errors}, C^*_x = Attr(\tau_i),
    \end{aligned}
\end{equation}
where $\tau_i$ is produced when target agent $A$ executes $tc_i$. $C^{errors}$ denotes the set of capabilities that are erroneous (i.e., cannot be verified by the oracle).
$Attr()$ represents the function, which attributes the decisive error in the $\tau_i$ to the sole capability $C^*_x$.




\section{Approach}
\label{sec:approach}

Our proposed {\tool} contains three modules, whose overview is shown in Figure \ref{fig:overview}.
First, we design a module of \textit{Adaptive Test Case Generation}, which utilizes an adaptive generation mechanism to create challenging task instructions, as test cases for Vision-and-Language Navigation (VLN) tasks.
The second module is \textit{Capability Oracles Construction}. 
It automatically constructs oracles to determine if there are errors in the capabilities, by obtaining the expected output through prior knowledge and the execution state of the target agent and defining metrics to measure the deviation between the actual and expected output.
Finally, in the \textit{Capability-Oriented and Failure-Oriented Feedback} module, we introduce a feedback mechanism. 
It aims to attribute failures to specific capabilities based on oracles, thereby calculating capability-oriented feedback, which is then combined with failure-oriented feedback to guide the iterative generation of test cases.

\subsection{Adaptive Test Case Generation}
\label{sec:case-generation}

Following the classical search-based fuzzing methodology \cite{fuzz,MoDitector}, {\tool} maintains a seed corpus where each seed corresponds to a test case, with a feedback score. A higher feedback score indicates a higher likelihood of uncovering capability-oriented failures.
The feedback score will be further elaborated in Section \ref{sec:feedback}.
In each iteration, adaptive generation first selects seeds and then applies mutation to the selected seeds to generate new test cases.

Our scene assets are sourced from the HM3D dataset \cite{HM3D}, which provides a rich collection of 3D panoramic scenes, including semantic annotations of rooms and objects (denoted as $A_r$ and $A_o$, respectively), forming the foundation for test case creation. Using these assets as the initial resource pool, custom-designed prompts are fed into a large language model (LLM), which combines room and object annotation to generate the task instruction: $I = \mathbb{L}(a_r, a_o)$, where $a_r \in A_r$ and $a_o \in A_o$. We treat task instructions as test cases for target agents. The prompt is shown in Appendix \ref{sec:app-prompt}.

{\tool} first initializes a batch of test cases as the initial seed corpus $SC$.
During each iteration of seed updates, we first calculate the selection probability for all seeds and then select a candidate based on these probabilities.
The selection probability for each case seed $p_{cs_i}$ is:
\begin{equation}
    \label{equation:selection-probability}
    p_{cs_i} = \frac{max(F_{cs_i}, 0)}{\sum^N_{i=1}F_{cs_i}},
\end{equation}
where the denominator is the sum of all feedback scores. The higher the feedback score, the higher the probability of the seed being selected.

After selecting a seed $s$, {\tool} aims to balance \emph{failure preservation} and \emph{search-space exploration} during mutation. 
To this end, we design two instruction mutation operators with different intensities: \emph{mild mutation}, which makes small semantic changes to preserve failure-relevant part, and \emph{aggressive mutation}, which introduces larger changes to diversify trajectories and expose additional weaknesses. 
Moreover, {\tool} adopts an adaptive strategy to select a mutation operator based on feedback score. 
Detailed mutation operator, selection strategy, and algorithm are provided in Appendix~\ref{sec:app-case-generation} and \ref{sec:app-prompt}.


\subsection{Capability Oracles Construction}
\label{sec:oracles-construction}

Achieving capability-oriented testing requires constructing capability oracles. To address it, we propose an automated oracle-construction mechanism that derives the expected output of each capability, and then compares it with the agent’s actual outputs to calculate metrics for identifying deficiencies.

We leverage expert models provided in the simulation environment to obtain the true task execution traces and related information.
Specifically, the navigation expert has access to the global map and uses greedy pathfinding to produce a (near-)optimal route \cite{Habitat3, Habitat}. For perception, we use the RAM image annotation model \cite{RAM} to provide expected detections and semantic annotations (e.g., object labels and bounding boxes) for the current view. When such privileged information is unavailable, similar supervision can be obtained via shortest path algorithms and manually annotated data.

\noindent{\textbf{Perception Oracle}}.
We use the weighted Intersection over Union (IoU) \cite{iou} to measure the error in perception, as it is a widely recognized metric.
The error is measured by:
\begin{equation}
    \label{equation:perception-oracle}
    \epsilon^p_t=\frac{1}{N} (\sum^N_{n=1} || VA_{t,n} - VA^{gt}_{t,n} ||_{\mathbb{L}} - \frac{|P_{t,n}\cap P^{gt}_{t,n}|}{|P_{t,n}\cup P^{gt}_{t,n}|}),
\end{equation}
where $t$ represents the $t$-th time step in the trajectory, $N$ represents the number of objects detected in the current view. $VA_{t,n}$ and $VA^{gt}_{t,n}$ denote the visual annotations of the target agent and the expected output, respectively.
The first term in the parentheses represents the similarity of visual annotations, and the second term corresponds to the IoU metric.
The term $|| VA_{t,n} - VA^{gt}_{t,n} ||_{\mathbb{L}}$ is provided by the LLM, where the prompt we use is shown in Appendix \ref{sec:app-prompt}. 
$P_{t,n}$ and $P^{gt}_{t,n}$ correspond to the detected bounding boxes obtained by the target agent's perception and the expected output of perception from expert model, respectively.

\noindent{\textbf{Memory Oracle}}.
We define a memory oracle that measures the accuracy of the ability to recall past information.
{\tool} first records the visual annotations information in the historical route and organizes them in terms of the time step, which is treated as the expected output of the memory content.
Then {\tool} calculates time step-oriented semantic similarity between current memory content of agent and visual annotations in the expected output to quantify errors of the memory.
\begin{equation}
    \label{equation:memory-oracle}
    \epsilon^m_t = 1-||M_t-VA^{gt}_{1,\dots,t-1}||_{\mathbb{L}},
\end{equation}
where $M_t$ represents the memory description from the target agent at time step $t$. 
$VA^{gt}_{1,\dots,t-1}$ denotes the sequence of visual annotations in the historical route before time step $t$, and their similarity is evaluated by the LLM (restricted between 0 and 1). The designed prompt is shown in Appendix \ref{sec:app-prompt}. 

\begin{figure*}[tbp]
    \centering
    \includegraphics[width=\textwidth]{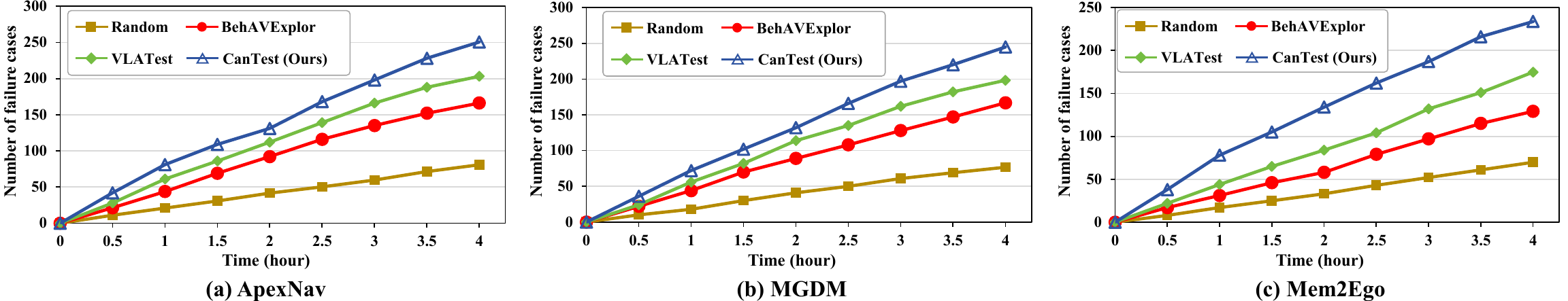}
    \caption{
    The comparison between {\tool} and the baselines on the number of failure cases for all target models.
    }
    \label{fig:rq1}
\end{figure*}

\begin{table*}[tbp]
\centering
\caption{Comparison of the number of failure cases toward each capability.
+\textit{\textbf{OA}} represents the integration of the capability \textbf{\textit{O}}racle and \textbf{\textit{A}}ttribution mechanism from our proposed {\tool}, as a baseline plug-in.
}
\label{tab:Failure_Attribution}
\resizebox{\linewidth}{!}{
\begin{tabular}{c|ccc|ccc|ccc|ccc}
\toprule
\multirow{2}{*}{\textbf{\makecell{Approach}}}  & \multicolumn{3}{c|}{\textbf{Perception}}    & \multicolumn{3}{c|}{\textbf{Memory}}   & \multicolumn{3}{c|}{\textbf{Planning}}    & \multicolumn{3}{c}{\textbf{Decision}}                                            \\
\cmidrule(r){2-13}
                                & \textbf{ApexNav}    & \textbf{MGDM}     & \textbf{Mem2Ego}            & \textbf{ApexNav}        & \textbf{MGDM}   & \textbf{Mem2Ego}                & \textbf{ApexNav}        & \textbf{MGDM}   & \textbf{Mem2Ego}            & \textbf{ApexNav}        & \textbf{MGDM}   & \textbf{Mem2Ego} \\
\midrule
{\textbf{Random + \textit{OA}}}               & 20.4	& 22.8	& 21.1	& 16.4	& 15.7	& 11.1	& 19.2	& 18.1	& 16.9	& 24.8	& 20.1	& 21.7                           \\
{\textbf{BehAVExplor + \textit{OA}}}          & 40.9	& 43.1	& 37.1	& 46.4	& 44.2	& 20.5	& 35.6	& 37.4	& 33.1	& 43.2	& 42.1	& 38.4                          \\
{\textbf{VLATest + \textit{OA}}}              & 51.8	& 56.4	& 50.1	& 54.1	& 50.4	& 26.9	& 45.1	& 41.1	& 39.7	& 52.1	& 50.4	& 58.1                         \\
\rowcolor{gray!20}
{\textbf{{\tool} (Ours)}}                 & \textbf{72.2}	& \textbf{74.7}	& \textbf{61.4}	& \textbf{66.3}	& \textbf{56.1}	& \textbf{42.8}	& \textbf{52.5}	& \textbf{49.3}	& \textbf{66.1}	& \textbf{59.5}	& \textbf{64.7}& \textbf{63.4}                           \\ 
\bottomrule
\end{tabular}
}
\end{table*}

\noindent{\textbf{Planning Oracle}}.
The error of planning capability can be measured by evaluating the similarity between the trajectory generated by target agent's planned route and ground truth trajectory provided by the expert model.
Since the planning process involves spatial and temporal reasoning, we adopt the normalized Dynamic Time Warping (nDTW) \cite{ndtw} to assess the alignment between the two trajectories.
nDTW is widely used for path similarity evaluation, as it accounts for both spatial proximity and temporal alignment.
The error of the planning capability is measured by:
\begin{equation}
    \label{equation:planning-oracle}
    \epsilon^{pl}_t = 1 - \text{nDTW}(\tau^{pl}_t, \tau^{gt}_{t,\dots,n}),
\end{equation}
where $\tau^{pl}_t$ represents route planned by target agent at time step $t$. $\tau^{gt}_{t,\dots,n}$ denotes the ground truth trajectory sequence from time step $t$ onwards.

\noindent{\textbf{Decision Oracle}}.
The decision commands are directly applied to the target agent and should follow the planned action provided by the upstream planning.
Since the set of available actions is finite and fixed, we directly compare the differences between the actual decision actions and the planned actions in the trajectory: 
\begin{equation}
    \label{equation:decision-oracle}
    \epsilon^{d}_t = 1 - ||D_t - D^{pl}_t||,
\end{equation}
where $D_t$ represents the action chosen by the target agent at time step $t$.
$D^{pl}_t$ is the planned action from the previous planned trajectory $\tau^{pl}_{t-1}$, which serves as the expected output for the decision.

\subsection{Capability-Oriented and Failure-Oriented Feedback}
\label{sec:feedback}
After obtaining the task instruction (i.e., test case) and capability oracles, {\tool} lets the target agent perform the task.
The resulting trajectory $\tau = \{s_1, s_2, \dots, s_T\}$ is a time series, where the state $s_t$ at each time step contains both the agent's own information (e.g., current position) and information about the surrounding environment (e.g., nearby objects).
{\tool} utilizes capability oracles to determine whether the current test case fails due to errors of the specific capability.
Then, the failure-oriented and capability-oriented feedback scores are computed to simultaneously guide adaptive generation of high-value test cases.
The algorithm of feedback calculation is shown in the Appendix \ref{sec:app-feedback}.

\subsubsection{\textbf{Capability-Oriented Failure Attribution}}
To attribute a task failure to specific capabilities, {\tool} identifies which capability errors are truly responsible for the failure along a long trajectory. 
First, {\tool} detects \emph{capability errors} using the capability oracles (Sec.~\ref{sec:oracles-construction}) for $\{C_p, C_m, C_{pl}, C_d\}$. 
Because not every detected error necessarily causes the final failure \cite{DecisiveError}, we further determine \emph{failure-inducing errors} via counterfactual reasoning \cite{EMAI}: for each detected error at time $t$ in capability $C_x$, we intervene by replacing the agent output with the oracle output and roll out the remainder of the trajectory under this correction. 
If this intervention turns the original failed trajectory into a success, the error is deemed failure-inducing. 
When multiple failure-inducing errors exist, we attribute the failure to the earliest one as the \emph{failure-source error}, and refer to its corresponding capability as the \emph{failure-source capability}. 
Details of the intervention operator, indicator function, and the earliest-error rule are provided in Appendix~\ref{sec:app-Failure-Attribution}.

\begin{figure*}[tbp]
    \centering
    \includegraphics[width=\textwidth]{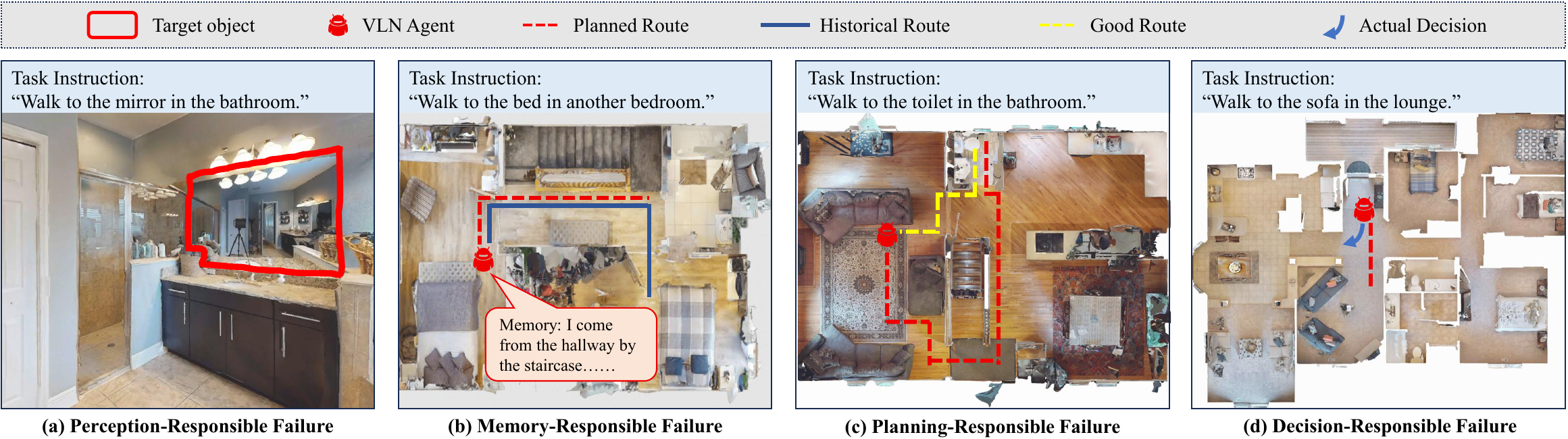}
    \caption{
    Examples of Failure Due to Different Capabilities.
    }
    \label{fig:example}
\end{figure*}

\subsubsection{\textbf{Feedback of Test Case}}
For the test case seed, the feedback score consists of two components: (1) \textit{Failure-Oriented Feedback} and (2) \textit{Capability-Oriented Feedback}.
Failure-oriented feedback aims to guide the generation of the case seed that is more likely to result in failures.
Capability-oriented feedback aims to guide the generation of the case seed that is more likely to result in capability-oriented failures.

\textbf{(1) Failure-Oriented Feedback}.
Based on whether the task is successful, failure-oriented feedback $F^{f}$ is defined as $0$ (failure) or $1$ (success).

\textbf{(2) Capability-Oriented Feedback}.
We identify the failure-source capability and the failure-source time step, i.e., $(C^*_x, t^*)$, as described above.
For capability $C^*_x$, the corresponding error value $\epsilon^x_{t^*}$ is calculated by the capability oracles.
The error values for each capability are normalized to the range $[0,1]$ and are used as the capability-oriented feedback, denoted as $F^{c} = Norm(\epsilon^x_{t^*})$.

The final feedback score $F_{cs}$ of the case seed integrates both failure-oriented and capability-oriented aspects: $F_{cs} = F^{f} + \lambda^{C_x} F^{c}$.
$\lambda^{C_x}$ is an adaptive parameter used to control the weight of capability-oriented feedback.
For failure-source capability $C^*_x$, if the number of failure cases caused by $C_x$ in the already identified capability-oriented failure case set is relatively high, a lower $\lambda^{C_x}$ is used to reduce the search focus on this capability. 
It helps ensure comprehensive testing across all capabilities. $\lambda^{C_x}$ is defined as:
\begin{equation}
   \label{equation:lambdac}
   \lambda^{C_x} = {\overline{N^{C_y}}} /{N^{C_x}},\ C_y \in \{C_p, C_m, C_{pl}, C_d\},
\end{equation}
where $\overline{N^{C_y}}$ denotes the average of capability-oriented failure cases in the failure case set. This score is used to guide adaptive test case generation.

\section{Evaluation}
\label{sec:evaluation}

\begin{table*}[tbp]
\centering
\caption{The results of repairing failure cases using oracles.
\#Fail represents the number of discovered failure test cases. \#Repa and \%Repa respectively represent the number and proportion of test cases successfully repaired.
}
\label{tab:repairing}
\resizebox{\linewidth}{!}{
\begin{tabular}{c|ccc|ccc|ccc|ccc}
\toprule
\multirow{2}{*}{\textbf{\makecell{Model}}}  & \multicolumn{3}{c|}{\textbf{Perception}}    & \multicolumn{3}{c|}{\textbf{Memory}}   & \multicolumn{3}{c|}{\textbf{Planning}}    & \multicolumn{3}{c}{\textbf{Decision}}                                            \\
\cmidrule(r){2-13}
                                & \textbf{\#Fail}    & \textbf{\#Repa}     & \textbf{\%Repa}       & \textbf{\#Fail}    & \textbf{\#Repa}     & \textbf{\%Repa}    & \textbf{\#Fail}    & \textbf{\#Repa}     & \textbf{\%Repa}    & \textbf{\#Fail}    & \textbf{\#Repa}     & \textbf{\%Repa}\\
\midrule
{\textbf{ApexNav}}               & 72.2	& 60.9	& 84.35\%	& 66.3	& 53.9	& 81.30\%	& 52.5	& 45.7	& 87.05\%	& 59.5	& 56.6	& 95.13\%                           \\
{\textbf{MGDM}}          & 74.7	& 62.4	& 83.53\%	& 56.1	& 46.2	& 82.35\%	& 49.3	& 42.6	& 86.41\%	& 64.7	& 61.4	& 94.90\%                          \\
{\textbf{Mem2Ego}}              & 61.4	& 52.7	& 85.83\%	& 42.8	& 35.8	& 83.64\%	& 66.1	& 59.3	& 89.71\%	& 63.4	& 61.3	& 96.69\%                         \\ 
\bottomrule
\end{tabular}
}
\end{table*}
\begin{figure*}[tbp]
    \centering
    \includegraphics[width=\textwidth]{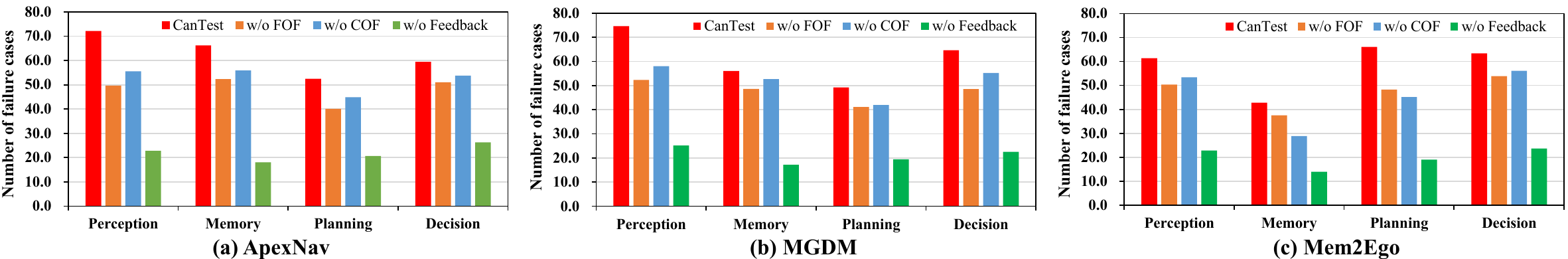}
    \caption{
    The results of the ablation study for feedback.
    }
    \label{fig:Feedback}
\end{figure*}


\noindent\textbf{Experimental Environment}.
We conduct experiments in Habitat3 \cite{Habitat3}, which provides a 3D scene environment for VLN.
We select three advanced VLN models as the target models to be tested: ApexNav \cite{apexnav}, MGDM \cite{MGDM}, and Mem2Ego \cite{mem2ego}.
Their details are shown in Appendix \ref{sec:app-Environment}.

\noindent\textbf{Baselines}.
We compare {\tool} with two commonly used techniques and one SOTA technique.
(1) \textbf{Random} is an approach for randomly generating test cases, which we implement following previous work \cite{demo2test, MDPFuzz}. 
(2) \textbf{BehAVExplor} \cite{BehAVExplor} is a behavior-guided fuzzing technique, which is commonly used to generate diverse test cases. 
(3) \textbf{VLATest} \cite{VLATest} is the SOTA fuzzing framework designed to generate robotic manipulation scenes for testing Vision-Language-Action (VLA) models. 

\subsection{Advantage of {\tool}}

\subsubsection{\textbf{Number of Failures}}
We present the trend of the number of discovered failure cases across the entire experimental duration of {\tool} and the baselines in Figure \ref{fig:rq1}. The results clearly demonstrate that, throughout each time interval, {\tool} consistently outperforms the baselines. 
The improvement in the final number of discovered failure cases relative to the best-performing baseline (VLATest) ranges from 23.34\% to 33.70\%.
This superior performance can be attributed primarily to the unique function of {\tool} to construct oracle for various underlying capabilities.
These oracles play a pivotal role in identifying the source errors in failure test cases, thereby providing more specific guidance that discovers failure cases more targeted and effective.

The feedback scores produced by this process are instrumental in guiding the iterative generation of test cases, ensuring that the focus remains on capabilities where the model struggles the most.
In contrast, BehAVExplor and VLATest rely solely on feedback from system-level outcomes to direct their case generation, neglecting the capability evaluation.
This limitation results in a less nuanced understanding of the specific weaknesses of the agent, hence leading to their relatively lower performance in uncovering diverse failure cases.

\subsubsection{\textbf{Number of Capability-Oriented Failures}}

Our proposed {\tool} can generate failure cases that can be attributed to specific capabilities. Moreover, its components of capability oracle and attribution mechanism can be used as a plug-in (denoted as + \textit{OA}) to extend baseline methods, thereby enabling them to perform capability-oriented failure attribution as well.
Based on this, we further compare the number of capability-oriented failures identified by our proposed {\tool} against those identified by the augmented baselines.
Table~\ref{tab:Failure_Attribution} presents the results across different capabilities.

The results indicate that {\tool} can discover the most failure cases for all capabilities. This reason is that {\tool} promotes comprehensive exploration of different capabilities through adaptive weighting when calculating the feedback scores to guide test case generation. Specifically, across the four capabilities, {\tool} significantly outperforms the baselines. This comprehensive coverage of capabilities ensures that a diverse range of failures can be identified during testing, providing more helpful failure analysis for agent enhancement.


\noindent\textbf{Examples of Capability-Oriented Failures}. 
To further demonstrate the practicality of {\tool}, we manually analyze the failure test cases attributed to the four capabilities it discovers, with representative examples illustrated in Figure~\ref{fig:example}. Based on these interpretable examples, we provide actionable enhancement suggestions for improving the agent. The details are presented in Appendix \ref{sec:app-example}.

\subsection{Fidelity of Capability Oracles}
\label{sec:Fidelity}
To evaluate the fidelity of automatically constructed capability oracles by {\tool}, we utilized the oracles introduced in Section \ref{sec:oracles-construction} to provide expected output for each capability.
For the failure cases identified by {\tool}, we repaired these cases by replacing the outputs of the erroneous capabilities with the expected output. If the oracles are accurate, the test cases should be easily repaired.

The results of repairing failure cases are shown in Table \ref{tab:repairing}. For each failure-responsible capability, a relatively high repair rate is achieved, i.e., repair rates ranging from 81.30\% to 96.69\%.
This indicates that the expected output used by the oracles constructed by {\tool} is reliable and accurate.
Specifically, the repair rates for perception, memory, and planning, which are upstream capabilities, are lower compared to decision, which is a downstream capability.
This indicates that upstream errors are often more complex, as they can propagate and impair subsequent decision execution.
Across different capabilities, however, all models achieved high repair effectiveness, with repair rates exceeding 80\% when using the oracles, thereby validating the effectiveness of the automatically constructed capability oracles in {\tool}.

\subsection{Ablation: Feedback Effectiveness}
To validate the effectiveness of feedback, we compared {\tool} with three variants: (1) without failure-oriented feedback, using only capability-oriented feedback (w/o FOF), (2) without capability-oriented feedback, using only failure-oriented feedback (w/o COF), and (3) without any feedback (w/o Feedback). Figure \ref{fig:Feedback} displays the comparison of results between {\tool} and the different variants. In all cases, {\tool} outperformed the other variants, indicating that any type of feedback contributes to discovering more failure cases.
Furthermore, the performance difference between w/o FOF and w/o COF is minimal, which suggests that failure-oriented and capability-oriented feedback are nearly equally important.


\section{Conclusion}
\label{sec:conclusion}



This paper presents {\tool}, an automated framework for capability-oriented testing of embodied agents. {\tool} generates capability-oriented test cases and constructs independent oracles for key capabilities, enabling systematic detection of capability errors by comparing agent outputs against expected outputs. To handle long-horizon interactions, {\tool} further introduces a failure attribution mechanism to identify failure-source errors along the trajectory. Experiments demonstrate that {\tool} discovers more failure cases than strong baselines, achieves the best capability-level attribution performance, and attains a high repair rate, validating the fidelity of the proposed oracles. Ablation studies further confirm the effectiveness of our failure- and capability-oriented feedback signals.

\section*{Limitations}

Despite its effectiveness, {\tool} has several limitations that remain to be addressed in future work.

\noindent\textbf{Expert Model}. 
A key assumption in {\tool} is the availability of an ``expert'' model to construct capability-specific oracles, e.g., (i) an optimal or near-optimal route for navigation/planning, and (ii) semantic grounding for perception (e.g., room/object labels). 
In settings where such an expert is not directly provided, obtaining reliable supervision may require additional prior knowledge or effort. For instance, in structured maps or partially known layouts, shortest-path planning (e.g., over a reconstructed graph) can serve as a proxy expert for near-optimal routes. Similarly, perception supervision can be approximated through extra annotation effort, such as human-labeled semantic descriptions of scenes or objects.

\noindent\textbf{Simulation Environment}. 
Our current evaluation is conducted in simulation, and directly transferring our oracle design to the physical world is non-trivial.
We view {\tool} as a diagnostic framework whose core idea—capability-oriented failure attribution via oracle interventions—remains applicable, but whose oracle implementations must be adapted for sim2real. In future work, we will explore (i) leveraging human-in-the-loop supervision for sparse but high-value annotations (e.g., validating key perception entities or waypoint-level plans) to calibrate oracle thresholds; and (ii) training learned surrogate oracles from real logs, where ``expert'' signals can be distilled from demonstrations, corrective feedback, or safety monitors. We expect that combining weak experts with uncertainty-aware oracle thresholds and selective human verification can enable {\tool} to provide useful, capability-level failure attribution under a realistic environment.

\section*{Acknowledgments}
This work was supported by the National Natural Science Foundation of China Grant No. 62232016, 
Basic Research Program of ISCAS Grant No. ISCAS-JCZD-202405 and No. ISCAS-JCZD-202304, Major Program of ISCAS Grant No. ISCAS-ZD-202401 and No. ISCAS-ZD-202302, Innovation Team 2024 ISCAS (No. 2024-66),
the National Research Foundation, Singapore, and the Cyber Security Agency under its National Cybersecurity R\&D Programme (NCRP25-P04-TAICeN). Any opinions, findings and conclusions or recommendations expressed in this material are those of the author(s) and do not reflect the views of National Research Foundation, Singapore and Cyber Security Agency of Singapore.


\bibliography{refer}

\appendix
\newpage
\section{Appendix}

\subsection{Background}
\label{appx:background}

\subsubsection{Vision-Language Navigation}
In this section, we take the VLN agent as the research target, which is an intelligent agent that operates in an unknown or partially known environment to perform navigation tasks based on specific instructions.
Specifically, the VLN agents aim to enable intelligent agents to perform navigation tasks in unknown environments based on natural language instructions \cite{vln_plan1}.
The agent receives a natural language instruction $I$ as input (e.g., "Walk from the kitchen to the living room, then enter the green door leading to the bedroom").
The instruction is represented as a sequence of words, denoted as: $I = \{w_1, w_2, w_3, \dots, w_L\}$,
where $w_i$ denotes the $i$-th word in the instruction, and $L$ is the total number of words. 

The agent perceives the environment from an egocentric (first-person) view and interacts with it through a series of discrete viewpoint changes or low-level actions (e.g., moving forward, turning left or right).
The agent's execution generates a trajectory $\tau = \{s_1, s_2, \dots, s_T\}$ formed as a time sequence, where $s_t$ represents the agent's state at the $t$-th time step.
The task may contain incomplete or ambiguous instructions, as well as dynamic elements and noisy visual inputs in the environment, presenting substantial challenges for navigation \cite{Fine-grained}.

\subsubsection{Failures of VLN Task}
In the VLN task, failures of the agent are typically defined as the inability of the agent to successfully complete the given navigation task.
Specifically, failures can be determined from the following aspects \cite{vln_state}:

\textbf{Target Position}: For each navigation task, after the agent executes the stop action, the task is considered successful if the distance between the agent's final position and the target position is within a certain threshold. The formula is as follows:
\begin{equation}
    \label{equation:final-position}
   Fail_{pos} = 
   \begin{cases} 
   True, & \text{if } d(p_{\text{final}}, p_{\text{target}}) > \delta, \\
   False, & \text{otherwise},
   \end{cases}
\end{equation}
where $p_{\text{final}}$ denotes the agent's final position, $p_{\text{target}}$ is the target position, $d(\cdot, \cdot)$ represents the distance function, and $\delta$ is the distance threshold.

\textbf{Step Limit}: If the number of steps taken by the agent exceeds the maximum step limit, the task is considered a failure.
The formula is as follows:
\begin{equation}
    \label{equation:step-limit}
   Fail_{step} = 
   \begin{cases} 
   True, & \text{if } t > T_{\text{max}}, \\
   False, & \text{otherwise},
   \end{cases}
\end{equation}
where $t$ is the number of steps taken by the agent, and $T_{\text{max}}$ is the maximum allowed number of steps for the task.

\subsubsection{\textbf{Capabilities of VLN Agent}}
To accomplish VLN tasks, agents usually require the following core capabilities \cite{vln1, vln2}:
\begin{itemize}
    \item \textbf{Perception}: The agent perceives the environment from an egocentric view, extracting visual features (e.g., objects, obstacles, spatial layout) \cite{vln_Perception}. Perception directly affects the agent's ability to link observations with language instructions (e.g., identifying a "green door").
    \item \textbf{Memory}: VLN tasks often require short-term and long-term memory to record previously visited locations and observed scenes, helping the agent avoid revisiting the same areas \cite{vln_memory}.
    \item \textbf{Planning}: Planning involves designing an efficient navigation trajectory to the target, considering both language instructions and the observed environment \cite{vln_plan2}. For example, the agent may need to infer a path to a "bedroom" through an intermediate room ("living room").
    \item \textbf{Decision}: Decision-making determines the next action (e.g., turn left, move forward) based on real-time perception, memory, and planned trajectories \cite{vln_plan3}.
\end{itemize}

These capabilities are interdependent during task execution. 
Perception provides the foundation for memory and planning, memory supports planning and decision by reducing redundant actions, and planning defines the trajectory that decision follows in real-time.
While these capabilities collectively enable VLN agents to navigate effectively, any deficiency in these capabilities can lead to task failures.

\subsection{Adaptive Test Case Generation}
\label{sec:app-case-generation}
We design an algorithm for adaptive case generation, described in Algorithm \ref{alg:case-generation}.
{\tool} first initializes a batch of test cases as the initial seed corpus $SC$ (line 1).
During each iteration of seed updates, we first calculate the selection probability for all seeds and then select a candidate based on these probabilities (line 2).
where the denominator is the sum of all feedback scores. The higher the feedback score, the higher the probability of the seed being selected.

After selecting seed $s$ as the candidate, we designed an adaptive mutation strategy that applies different mutation methods to candidates with varying levels of feedback scores.
Essentially, mutation involves modifying the objects or rooms in the task instruction so that the agent takes a different route, to make the route more likely to reveal failures.
We design prompts to apply the LLM to mutate the current instructions, and detailed prompts are available on our website.
Specifically, for task instructions, we design two levels of mutation as follows.

\begin{figure*}[tbp]
    \centering
    \includegraphics[width=\textwidth]{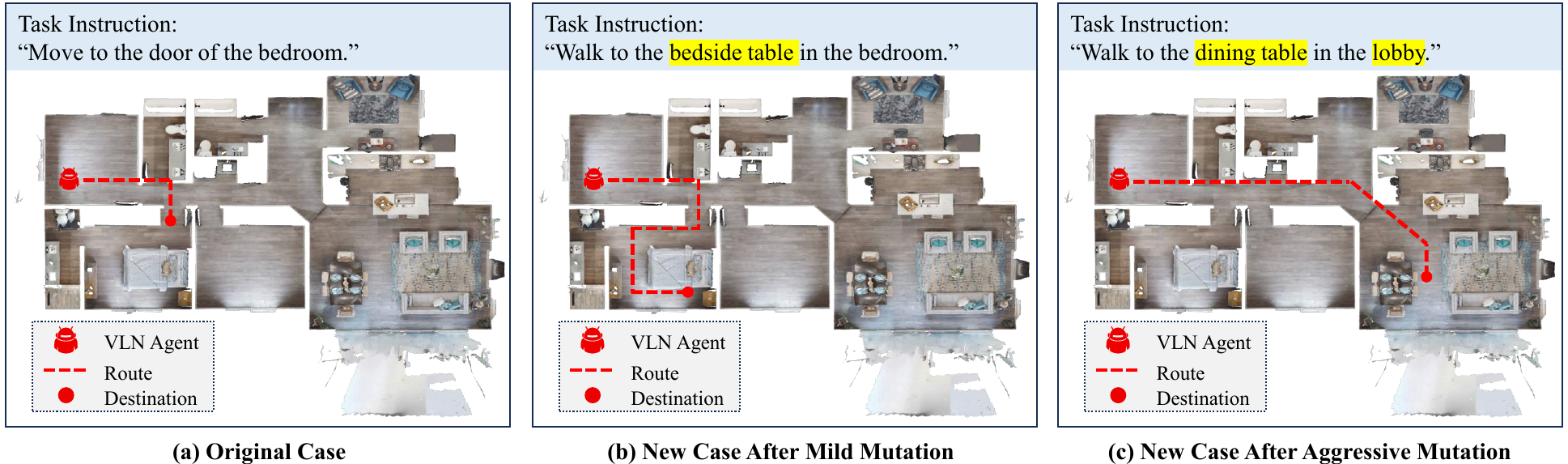}
    \caption{
    Illustration of the mild mutation and the Aggressive mutation.
    }
    \label{fig:mutation}
\end{figure*}

\textbf{Mild Mutation}: for test cases with higher feedback scores (indicating a higher probability of failure), we apply mild mutation to the task instruction. For example, as shown in Figures \ref{fig:mutation} (a) and (b), we make slight changes without altering the destination room (i.e., modifying the destination as another object within the room).

\textbf{Aggressive Mutation}: conversely, for those with lower feedback scores, as shown in Figures \ref{fig:mutation} (a) and (c), using the same room as the destination may not easily lead to failures. Therefore, we apply aggressive mutation to alter the target room itself, guiding the agent along a significantly different path to expose potential weaknesses.

To adaptively select the appropriate mutation strategy, we calculate the following mutation probabilities based on the feedback score:
\begin{equation}
    \label{equation:mutation-probability}
    p_{m} = \frac{F_{cs} - min(\mathbf{F})}{max(\mathbf{F}) - min(\mathbf{F})}, 
\end{equation}
where $\mathbf{F}$ = $\{F_{cs_1},\dots, F_{cs_n}\}$ and $p_m$ represents the degree to which the feedback score of the current seed ranks among all seeds.
A higher $p_m$ indicates that current seed is more outstanding within the seed corpus, i.e., it is more likely to fail due to certain capability deficiencies. 
Therefore, {\tool} tends to perform mild mutation to avoid excessive mutations that could lead to difficulty in revealing failure. 
Conversely, a lower $p_m$ leads to a tendency for aggressive mutation, expanding the search space and striving for greater difference before and after the mutation.

A higher $p_m$ indicates that the current seed is more outstanding within the seed library, i.e., it is more likely to fail due to certain capability deficiencies. 
Therefore, {\tool} tends to perform mild mutation to avoid excessive mutations that could lead to difficulty in revealing failure (lines 4-8). 
Conversely, a lower $p_m$ leads to a tendency for aggressive mutation, expanding the search space and striving for greater difference before and after the mutation (lines 9-13).

\begin{algorithm}[t]
    \caption{The Algorithm for Adaptive Test Case Generation.}
    \label{alg:case-generation}
    \KwIn{The semantic annotations $A_r$ and $A_o$ of environment assets.}
    \KwOut{The generated test case seed $s'$.}
    \textbf{Initialization:} Generate the initial seed corpus $SC$.                                                          \\
    Select candidate $s$ from $SC$. with probability $p_s$ according to the Equation \ref{equation:selection-probability};            \\
    Calculate the adaptive mutation probability $p_m$ according to the Equation \ref{equation:mutation-probability};        \\
    \If {random() < $p_m$}{
        Get the room annotation $a_r$ and object annotation $a_o$ from candidate $s$;                                        \\
        Sample another object: $a_o'$ = Sample($A_o$, $a_r$, $a_o$);                            \\
        $s'$ = MildMutation($s$, $a_r$, $a_o'$);                                            \\
    }
    \Else{
        Get the room annotation $a_r$ from candidate $s$;                                        \\
        Sample another room and object: $a_r'$, $a_o'$ = Sample($A_r$, $A_o$, $a_r$);                 \\
        $s'$ = AggressiveMutation($s$, $a_r'$, $a_o'$);                                         \\
    }
    \Return $s'$;
\end{algorithm}

\subsection{The Algorithm for Capability Failure Feedback}
\label{sec:app-feedback}
The workflow for capability and failure feedback is shown in the Algorithm \ref{alg:feedback}.
After obtaining the task instruction (i.e., test case) and capability oracles, {\tool} lets the target agent explore the environment based on the instruction to perform the task.
The resulting trajectory $\tau = \{s_1, s_2, \dots, s_T\}$ is a time series, where the state $s_t$ at each time step contains both the agent's own information (e.g., current position) and information about the surrounding environment (e.g., nearby objects).
The various capabilities generate their respective outputs based on $s_t$ (line 1).
{\tool} utilizes capability oracles to evaluate $\tau$, determining whether the current test case fails due to errors of the specific capability (line 2).
Then, the failure-oriented and capability-oriented feedback scores are computed to simultaneously guide adaptive generation of high-value test cases.

\begin{algorithm}[t]
    \caption{The Algorithm for Capability Failure Feedback.}
    \label{alg:feedback}
    \KwIn{The generated test case seed $s'$ and corresponding task instruction $I$, the oracles $\{TO_x\}$.}
    \KwOut{The feedback score $F_{cs}$ of the case seed.}
    
    Let target agent execute task according to instruction $I$ and get trajectory $\tau$;            \\
    Identify the capability error $(C_x, t)$ using the $\{TO_x\}$ introduced in Section \ref{sec:oracles-construction};        \\
    \If {result of $\tau$ $R(\tau)=Failure$ }{
        Replace the $(C_x, t)$ with the $\{TO_x\}$ to calculate indicator;         \\
        Identify the failure-inducing error $(C'_x, t')$ from $(C_x, t)$ using indicator;        \\
        Identify the failure-source error $(C^*_x, t^*)$ from $(C'_x, t')$;                                            \\
        Store $\tau$ to the failure case set;         \\
        Calculate the failure-oriented and capability-oriented feedback;                       \\
        Calculate the adaptive weight $\lambda^{C_x}$ for capability-oriented feedback;        \\
        \Return feedback score $F_{cs} = F^{f} + \lambda^{C_x} F^{c}$;
    }
    \Else{
        \Return feedback score $F_{cs}=0$;                                         \\
    }
\end{algorithm}

\subsection{The Details of Capability-Oriented Failure Attribution}
\label{sec:app-Failure-Attribution}
To attribute failures to specific capabilities, it is necessary to identify the source of failures.
Because capabilities interact across long task trajectories, not all capability errors necessarily lead to task failure.
To address this, {\tool} incorporates a failure attribution mechanism that first identifies capability errors via oracles, then determines which of these actually caused the failures (named as failure-inducing errors).
Among these, the earliest occurring error is designated as the failure-source error.
The capability corresponding to the error is referred to as failure-inducing or failure-source capability.

Specifically, for the capabilities of perception, memory, planning, and decision (denoted as $\{C_p, C_m, C_{pl}, C_d\}$), each capability error is evaluated by the oracle introduced in Section \ref{sec:oracles-construction}.
We use the tuple $(C_x, t)$ to represent an error, which means that the capability $C_x$ ($C_x \in \{C_p, C_m, C_{pl}, C_d\}$) of the target agent at time step $t$ produces outputs that violate the oracles.
A trajectory may contain multiple errors, but not all errors necessarily lead to an overall failure \cite{DecisiveError}.
We aim to identify the truly failure-inducing errors that ultimately lead to failure within a long chain of interdependent capabilities.

{\tool} attribute failures to specific capability errors, leveraging the concept of counterfactual causal reasoning \cite{EMAI}.
Specifically, suppose the trajectory $\tau$ is a failure, i.e., $R(\tau) = Failure$, as defined in Section \ref{sec:failure}.  
Consider the following scenario: if the error in capability $C_x$ at time $t$ is corrected, i.e., by replacing the original output with the capability oracle.  
The steps before time $t$ remain unchanged, while the actions after time $t$ are automatically updated accordingly.  
This process generates a modified trajectory:
\begin{equation}
   \label{equation:modified-trajectory}
   \tau^{(C_x, t)} = \mathbb{M}_{(C_x, t)}(TO_x, \tau),
\end{equation}
where $\mathbb{M}_{(C_x, t)}$ represents the modification to the output of capability $C_x$ at time step $t$.
$TO_x$ is the test oracle for this capability, which serves as the expected output for the capability output and is substituted into $\tau$. 
If the modified trajectory $\tau^{(C_x, t)}$ results in $R(\tau^{(C_x, t)}) = Success$, then the error $(C_x, t)$ is considered as a failure-inducing error.
We define a failure-inducing error indicator as a capability-time tuple $(C_x, t)$:
\begin{equation}
    \label{equation:decisive-indicator}
   \delta_\tau(C_x, t) = 
   \begin{cases} 
   1,     & \text{if } R(\tau^{(C_x, t)}) \text{ is } Success   \\
          &  \text{ and } R(\tau) \text{ is } Failure, \\
   
   0,     & \text{otherwise}.
   \end{cases}
\end{equation}

A failure-inducing error is then a tuple $(C'_x, t')$ with $\delta_\tau(C'_x, t') = 1$, where $C'_x$ denotes the \textit{failure-inducing capability} and $t'$ represents the corresponding failure-inducing time step.
Multiple failure-inducing errors may occur within a single trajectory.
To handle this, we attribute the failure to the earliest failure-inducing error, defined as:
\begin{align}
   \label{equation:earliest-tuple}
    \begin{aligned}
   & \mathbb{C}(\tau) = \{(C_x, t)|\delta_\tau(C_x, t) = 1\},     \\
   & (C^*_x, t^*) = \arg \min_{(C'_x, t') \in \mathbb{C}(\tau)}t, 
   \end{aligned}
\end{align}
where $C^*_x$ denotes the \textit{failure-source capability} and $t^*$ the corresponding failure-source time step.  
Accordingly, the failure of trajectory $\tau$ is attributed to the capability $C^*_x$ at time step $t^*$, and $\tau$ is stored in the failure case set.

\subsection{Details of Prompts}
\label{sec:app-prompt}
\begin{figure*}[tbp]
    \centering
    \includegraphics[width=\textwidth]{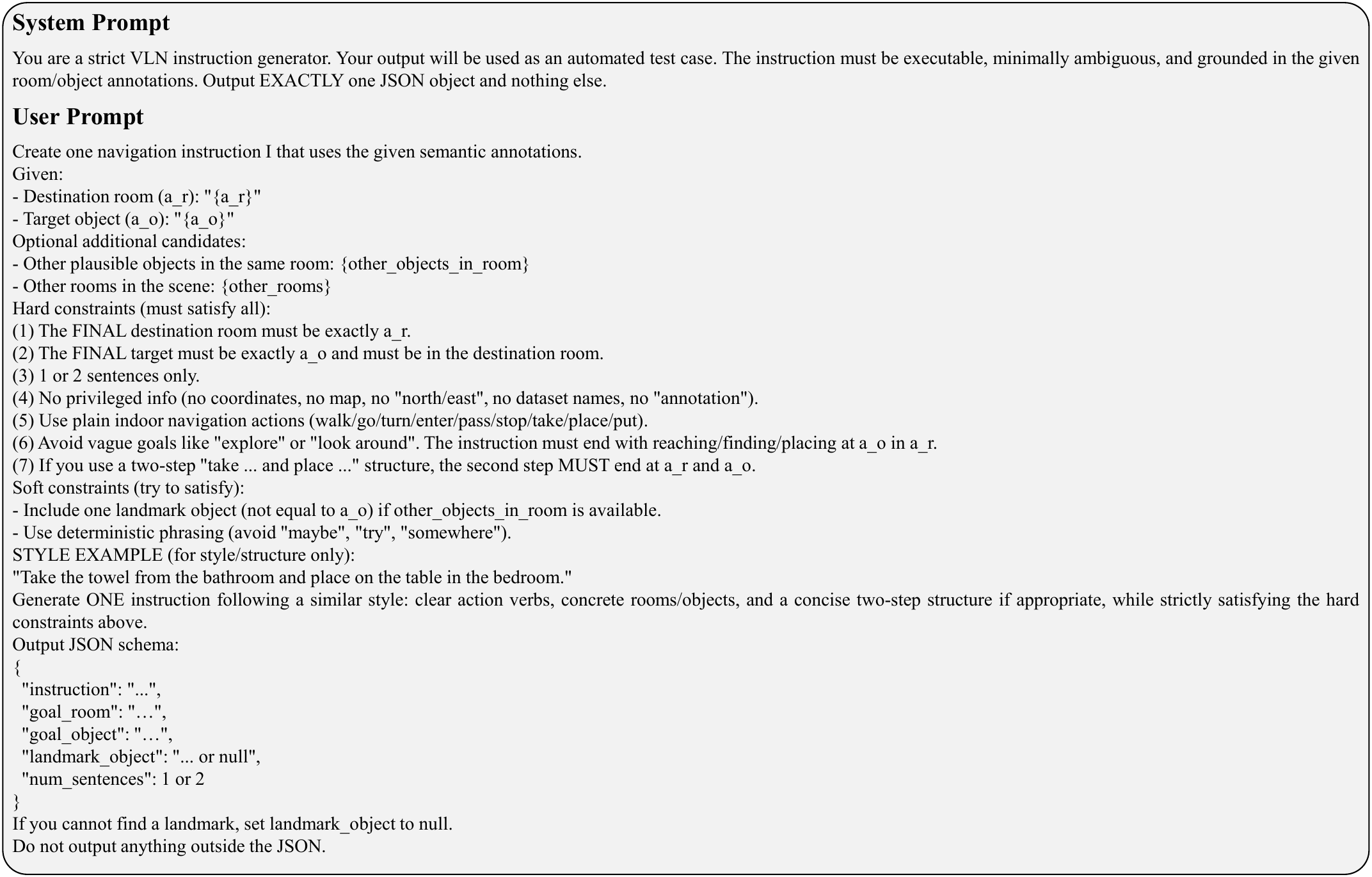}
    \caption{
    The prompt for task instruction generation.
    }
    \label{fig:generation_prompt}
\end{figure*}

\begin{figure*}[tbp]
    \centering
    \includegraphics[width=\textwidth]{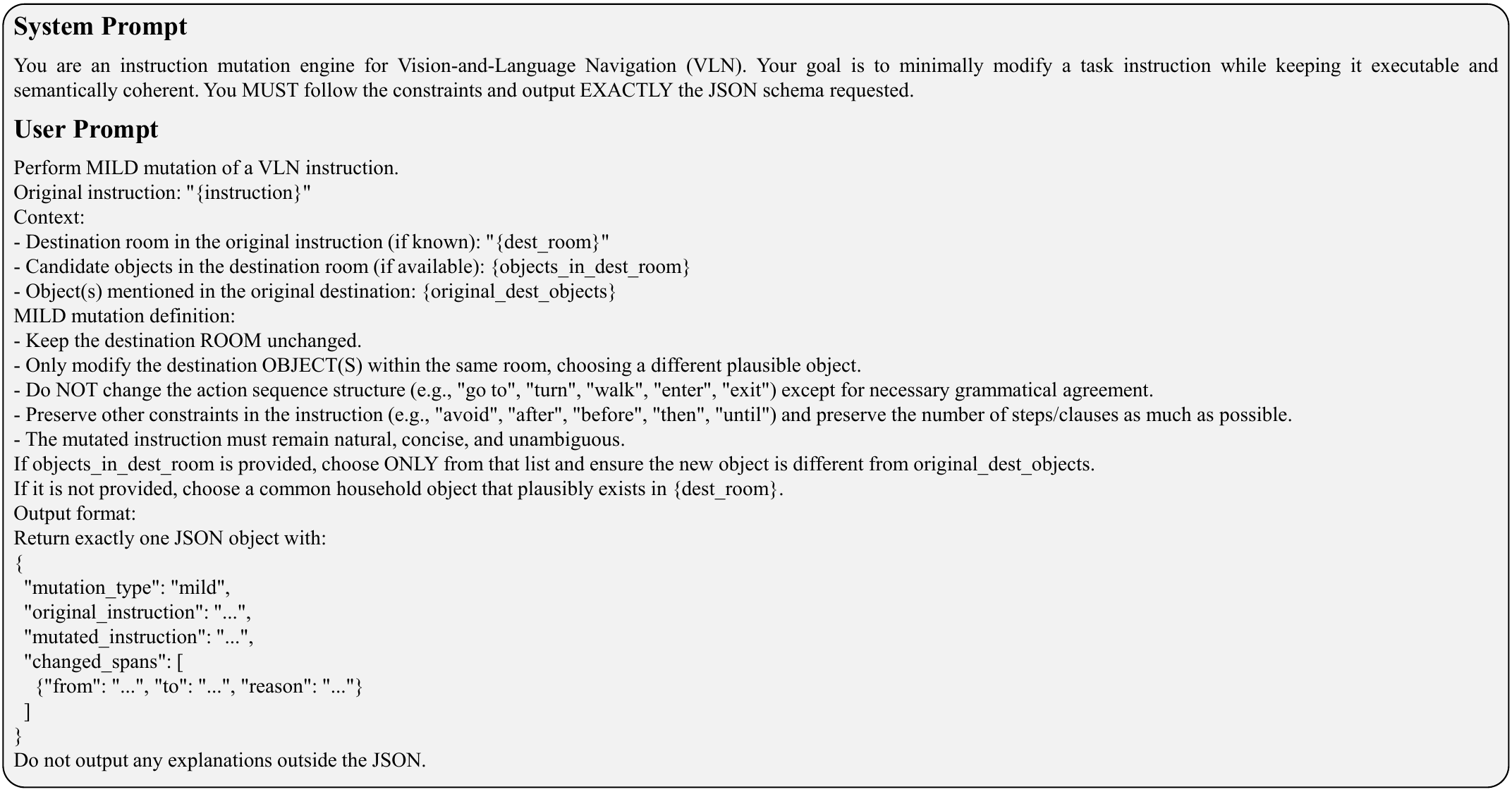}
    \caption{
    The prompt for mild mutation.
    }
    \label{fig:Mild_Mutation_Prompt}
\end{figure*}

\begin{figure*}[tbp]
    \centering
    \includegraphics[width=\textwidth]{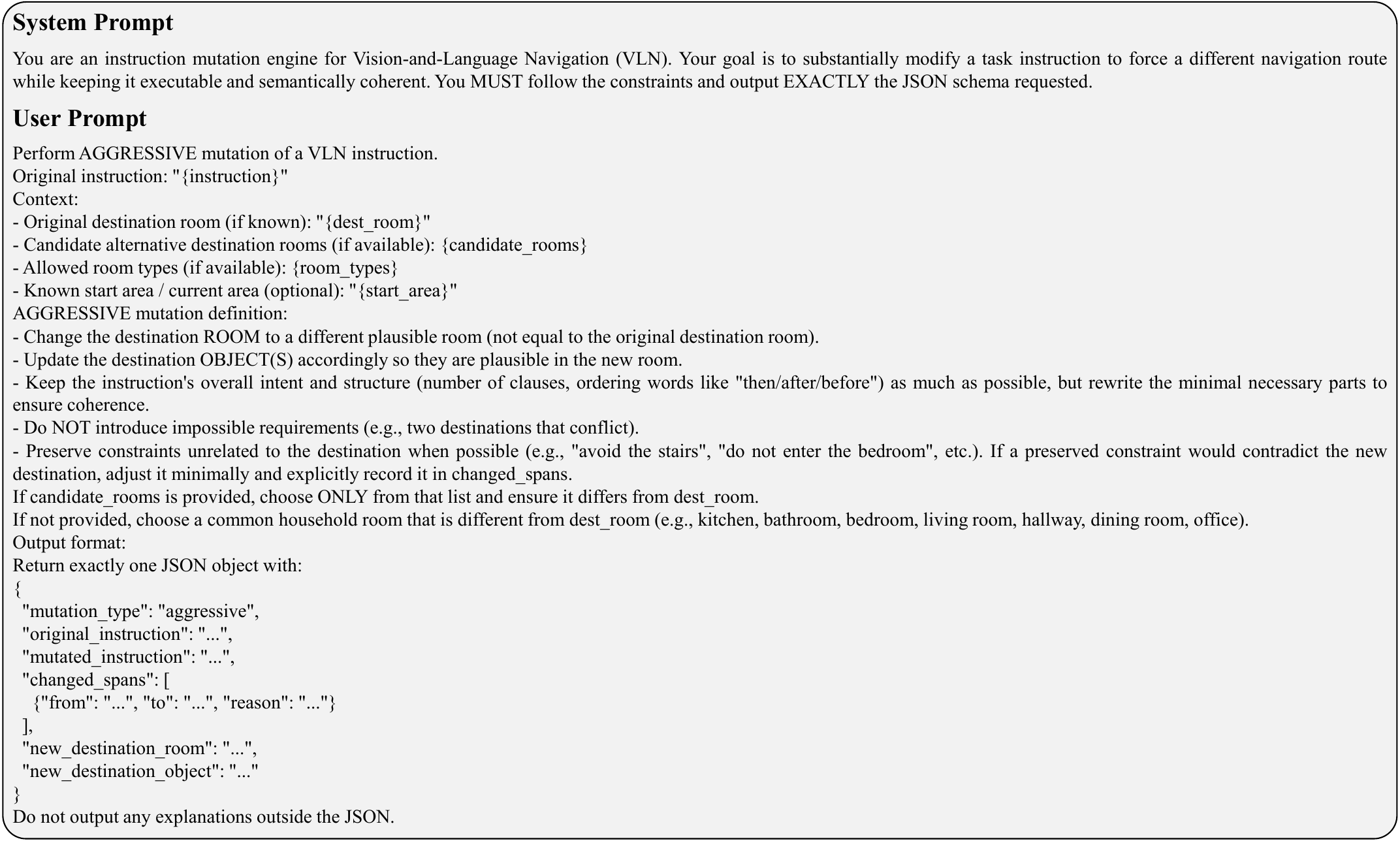}
    \caption{
    The prompt for aggressive mutation.
    }
    \label{fig:Aggressive_Mutation_Prompt}
\end{figure*}

\begin{figure*}[tbp]
    \centering
    \includegraphics[width=\textwidth]{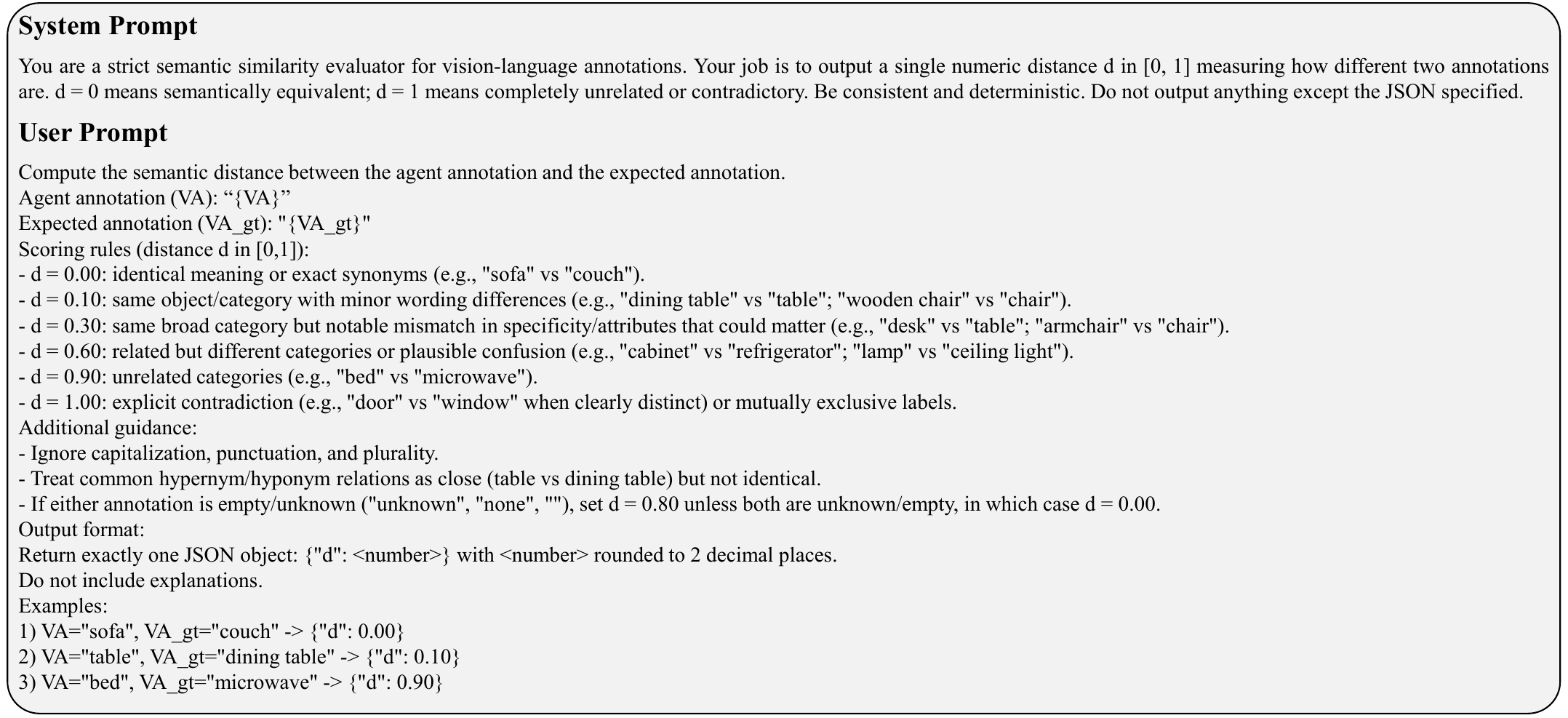}
    \caption{
    The prompt for perception oracle.
    }
    \label{fig:Perception_Prompt}
\end{figure*}

\begin{figure*}[tbp]
    \centering
    \includegraphics[width=\textwidth]{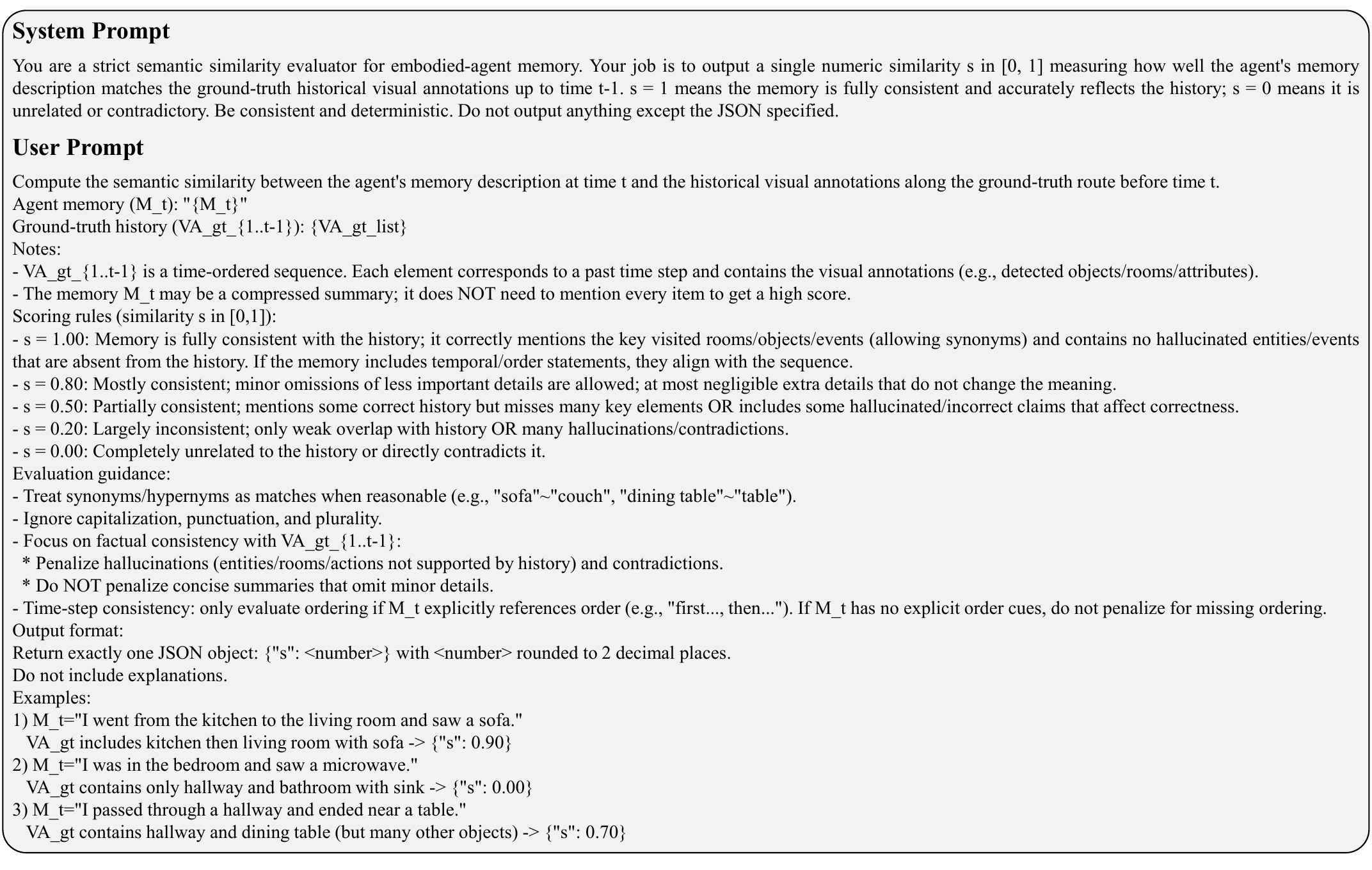}
    \caption{
    The prompt for memory oracle.
    }
    \label{fig:Memory_Prompt}
\end{figure*}

Figures~\ref{fig:generation_prompt}--\ref{fig:Memory_Prompt} summarize the prompts used in our framework. Specifically, we provide the templates for mild and aggressive mutation to generate candidate instructions, as well as the prompts for the perception and memory oracles that produce semantic observations and route/history information used by the planner.

\subsection{Experimental Settings}
\label{sec:app-Environment}
\subsubsection{Environment}
We take the VLN agent as the subject of our study, and conduct experiments in Habitat3 \cite{Habitat3}, a platform that provides a 3D scene environment for VLN.
At each time step, the agent receives RGB observations from three directions: forward, left, and right.
The agent is provided with atomic actions, including "Move Forward," "Turn Left," "Turn Right," and "Stop." When the agent executes the "Stop" action, the current task is considered complete.
The scene assets used in our experiments are primarily sourced from HM3D \cite{HM3D}, which includes 216 large-scale indoor 3D reconstructed scenes along with their semantic annotations.
We selected three advanced VLN models as the target models to be tested: ApexNav \cite{apexnav}, MGDM \cite{MGDM}, and Mem2Ego \cite{mem2ego}.

\subsubsection{Baseline}
The baseline VLATest defines four operators (Target object, Confounding objects, Lighting, and Camera) and generates test cases in a fuzzing-based framework by sampling/perturbing these factors. To adapt VLATest to VLN, we keep the same operator-driven generation paradigm and make task-specific adjustments: (1) Target object becomes the VLN’s destination landmark (goal-related object/room cue in the instruction); (2) Confounding objects become distractor landmarks that the agent should avoid/keep away from; (3) Lighting perturbs the perceived view via illumination-intensity changes; and (4) Camera perturbs observations via camera-rotation (viewpoint) shifts. The remaining components (e.g., object sampling strategy, perturbation magnitudes, and the fuzzing loop) follow the original VLATest framework.

\subsection{Failure Diversity}
\label{sec:app-Diversity}
To further examine whether the discovered failures are diverse rather than repeated instances of a few dominant patterns, we conduct an additional manual diversity analysis.
Specifically, we design a finer-grained failure taxonomy that goes beyond the four high-level capability sources used in the main paper.
The taxonomy contains eight failure types spanning perception, memory, planning, and decision-making, as summarized in Table~\ref{tab:failure_taxonomy_appendix}.

Under this taxonomy, we randomly sample 100 failure cases on ApexNav discovered by our method and another 100 cases discovered by the best baseline, and manually assign each case to one taxonomy category.
The results show that our method covers all eight categories, whereas the best baseline covers only six categories and does not discover failures of types \textbf{ME-2} (Temporal/Order Error) and \textbf{PL-2} (Looping).
This suggests that our method uncovers a broader range of failure modes, instead of repeatedly generating failures from the same pattern.

\begin{table*}[t]
\centering
\small
\caption{Finer-grained failure taxonomy used in the manual diversity analysis. Each failure type is defined by its characteristic evidence in the trajectory and its typical consequence.}
\label{tab:failure_taxonomy_appendix}
\resizebox{\linewidth}{!}{
\begin{tabular}{l|l|l|l}
\toprule
\textbf{Capability} & \textbf{Failure Type} & \textbf{Decision Evidence} & \textbf{Consequence} \\
\midrule
Perception & PE-1 Missed Detection & At step $t$, the target object or landmark is visible but not perceived. & Misses a critical entrance or landmark. \\

Perception & PE-2 Object Confusion & The target is perceived as a similar but different object or landmark. & Moves toward a similar but incorrect room or landmark. \\

Memory & ME-1 Forgetting & Previously visited paths or landmarks are not retained in memory. & Leads to redundant search or backtracking. \\

Memory & ME-2 Temporal/Order Error & The visitation order of landmarks is confused or misassigned. & Violates ordering constraints such as \textit{before}, \textit{after}, or \textit{pass-after}. \\

Planning & PL-1 Detour & Fails to choose a shorter route and instead follows a longer one. & Takes a long detour and reaches the step limit. \\

Planning & PL-2 Looping & Repeatedly passes the same landmarks or revisits the same area. & Falls into a loop until the step limit is reached. \\

Planning & PL-3 Constraint Violation & Fails to follow the instruction constraints. & The route violates constraints such as pass/avoid/order requirements. \\

Decision & DE-1 Inconsistent with Plan & The executed action deviates from the previously planned next step. & Makes an incorrect turn, e.g., at an intersection. \\
\bottomrule
\end{tabular}
}
\end{table*}

\subsection{Usefulness of Capability-Oriented Failures}
\label{sec:app-example}
\begin{enumerate}
    \item Perception-Oriented: as shown in Figure \ref{fig:example} (a), the agent is already positioned in front of the mirror as required by the instructions. However, due to insufficient perception capabilities, it cannot recognize the mirror, thus resulting in the agent moving away from the target object and being unable to complete the task instructions in the subsequent trajectory.
    
    Enhancement Suggestion:
    Improve the understanding of the environment by providing more context about the surrounding environment. For example, by integrating perception results from adjacent time steps, the Spatiotemporal continuity of context can be leveraged to reduce perception uncertainty.
    \item Memory-Oriented: as presented in Figure \ref{fig:example} (b), the instruction requires the agent to walk to another bedroom. However, after the agent traveled from one bedroom to another, it lost the long-term memory of originally coming from the first bedroom. This deficiency in memory subsequently cause the agent to leave the current bedroom again, leading to task failure.

    Enhancement Suggestion: Consider introducing a better integration mechanism of short-term and long-term memory, allowing the agent to retain necessary information while switching between different rooms. 
    
    \item Planning-Oriented: as illustrated in Figure \ref{fig:example} (c), the agent has different routes to reach the location specified in the task instructions. Due to poor planning, the agent did not choose a better route. The planned route is too long, ultimately causing the time steps to exceed the limit and resulting in failure. 
    
    Enhancement Suggestion: Improve the agent's path planning algorithms by allowing real-time analysis of the environment and dynamic adjustments to choose better routes. Additionally, set reasonable time budgets to ensure the agent considers time constraints during the planning process.
    \item Decision-Oriented: as shown in Figure \ref{fig:example} (d), the agent does not follow the planned route for decision-making. Due to its unwarranted autonomy in decision-making, it deviates from the planned route, ultimately leading to an improper decision that results in the inability to complete the task.

    Enhancement Suggestion: Limit the agent's autonomy in decision-making, and promptly assess whether adjustments are needed when decisions are inconsistent with the planned route.
\end{enumerate}

\end{document}